\documentclass[preprint,12pt]{elsarticle}

\usepackage[T1]{fontenc}
\usepackage[utf8]{inputenc}

\usepackage{amssymb}
\usepackage{multirow}
\usepackage{graphicx}
\usepackage{subcaption}
\usepackage{caption}
\usepackage{verbatim}
\usepackage{float}
\usepackage{amsmath}

\makeatletter
\def\ps@pprintTitle{%
  \let\@oddhead\@empty
  \let\@evenhead\@empty
  \let\@oddfoot\@empty
  \let\@evenfoot\@empty}
\makeatother

\usepackage[unicode]{hyperref}

\journal{Computers in Biology and Medicine}

\begin{document}
\begin{frontmatter}

\title{Enhanced Single-Cell RNA-seq Embedding through Gene Expression and Data-Driven Gene-Gene Interaction Integration\tnoteref{tn1}}
\tnotetext[tn1]{\textbf{Author Accepted Manuscript (AAM).} The Version of Record was published in 
\textit{Computers in Biology and Medicine}, Volume 188, April 2025, Article 109880.
Available at \url{https://doi.org/10.1016/j.compbiomed.2025.109880}.
Per Elsevier policy, this AAM is shared under a CC BY-NC-ND 4.0 license.}

\author[inst1]{Hojjat Torabi Goudarzi}
\affiliation[inst1]{organization={School of Electrical Engineering and Computer Science, Oregon State University},
            addressline={Address one}, city={Corvallis}, postcode={97331},
            state={Oregon}, country={United States}}

\author[inst2]{Maziyar Baran Pouyan}
\affiliation[inst2]{organization={Accenture Technology Labs},
            addressline={Address two}, city={San Francisco}, postcode={94105},
            state={California}, country={United States}}

\begin{abstract}
Single-cell RNA sequencing (scRNA-seq) provides unprecedented insights into cellular heterogeneity, enabling detailed analysis of complex biological systems at single-cell resolution. However, the high dimensionality and technical noise inherent in scRNA-seq data pose significant analytical challenges. While current embedding methods focus primarily on gene expression levels, they often overlook crucial gene-gene interactions that govern cellular identity and function. To address this limitation, we present a novel embedding approach that integrates both gene expression profiles and data-driven gene-gene interactions. Our method first constructs a Cell-Leaf Graph (CLG) using random forest models to capture regulatory relationships between genes, while simultaneously building a K-Nearest Neighbor Graph (KNNG) to represent expression similarities between cells. These graphs are then combined into an Enriched Cell-Leaf Graph (ECLG), which serves as input for a graph neural network to compute cell embeddings. By incorporating both expression levels and gene-gene interactions, our approach provides a more comprehensive representation of cellular states. Extensive evaluation across multiple datasets demonstrates that our method enhances the detection of rare cell populations and improves downstream analyses such as visualization, clustering, and trajectory inference. This integrated approach represents a significant advance in single-cell data analysis, offering a more complete framework for understanding cellular diversity and dynamics.
\end{abstract}


\end{frontmatter}

% ---------- AFTER front matter (will not be on page 1) ----------
\clearpage % optional: force Highlights to start on a new page
\section*{Highlights}
\begin{itemize}
\item A novel embedding method integrates gene expression with gene-gene interactions.
\item Introduces the Enriched Cell-Leaf Graph (ECLG) for improved cell representations.
\item Graph neural networks capture transcriptional states and regulatory relationships.
\item Enhances clustering, rare cell detection, and visualization over existing methods.
\item Optimized for large-scale single-cell RNA-seq with efficient feature extraction.
\end{itemize}

%% \linenumbers

%% main text
\clearpage 
\section{Introduction}
\label{sec:sample1}
    
scRNA-seq has fundamentally changed how we comprehend cellular heterogeneity, cell type diversity, and cellular dynamics in a wide range of biological processes and disease states by recording the transcriptomes of individual cells. By allowing transcriptome-wide analysis at the single-cell level, scRNA-seq enables researchers to understand the function and dynamics of individual cells within their native environments\cite{b1}.

The matter of dimensionality is one of the important challenges to be dealt with in scRNA-seq data analysis—each cell's gene expression profile can be depicted as a coordinate in a multi-dimensional space, where each dimension represents a particular gene\cite{b2}. 

The abundance of gene measurements in each cell results in scRNA-seq data that is usually characterized by high dimensionality. This brings difficulties when attempting to visualize and understand the data. To conquer this challenge, single-cell embedding techniques, have been employed to transform high-dimensional gene expression data into a lower-dimensional space \cite{b3}. Consequently, the utilization of single-cell embedding is vital for the analysis of scRNA-seq data. It aids in identifying different cell types through clustering, improves our comprehension of developmental and disease processes, and offers valuable insights for advancements in drug discovery and personalized medicine\cite{b4}.

Techniques such as Principal Component Analysis (PCA) \cite{b5}, t-distributed Stochastic Neighbor Embedding (t-SNE) \cite{b6}, and Uniform Manifold Approximation and Projection (UMAP) \cite{b7} have become standard tools in scRNA-seq data embedding for various analyses like clustering, visualization, and dimension reduction. In addition, researchers in \cite{b8} proposed a combination of multiple kernels (functions used to measure similarity or distance between cells) to learn a distance metric that's appropriate for single-cell RNA-seq data. Recently, advanced deep learning architectures have been developed for projecting scRNA expression data to a lower dimension \cite{b9, b10}. Additionally, recent advancements in deep learning-based technology have been successfully applied to healthcare problems, such as classifying fake news about COVID-19\cite{b11} and developing domain ontologies for Alzheimer’s disease\cite{b12}. However, Our approach expands on these methods by focusing on the integration of gene-gene interactions with gene expression profiles for single-cell RNA-seq.

Furthermore, a novel tree-based method for identifying a dissimilarity matrix between cells is introduced in \cite{b13}. This learned dissimilarity matrix can subsequently be fed into t-SNE or UMAP to derive a new embedding representation of the data. However, most of the mentioned techniques only pay attention to gene expression profiles, but do not consider the potential gene-gene interactions simultaneously. Gene-gene interaction, which is known as epistasis, has an important role in a cell's identity formation \cite{b14}. In biological systems, genes function within a network, interacting and impacting one another's behavior. However, present single-cell embedding methods fail to fully generalize on this crucial information while reducing the dimensionality of cell embeddings. As a result, the intricate interplay between genes remains inadequately utilized in current approaches.

In response to this important need, we have developed a novel method for single-cell embedding that integrates both gene expression profiles and gene-gene interaction information. Our method leverages the information contained in network-based analyses to capture the complex interplay between genes and integrates this information with conventional gene expression data to generate a more comprehensive representation of cellular states. Similarly, neural networks such as the Forward Only Counter Propagation Network have been applied to complex classification tasks, demonstrating the importance of capturing variable interactions in classification problems\cite{b15}.

Briefly, we extend the idea of GENIE3 (GEne Network Inference with Ensemble of Trees) which is a gene regulatory network extraction method proposed in \cite{b16}. GENIE3 has been shown to be effective in extracting gene regulatory networks from diverse datasets, including single-cell RNA-seq data, time-series microarray data, and ChIP-seq data \cite{b17}.

While previous studies have attempted to incorporate gene-gene interaction information into single-cell analysis \cite{b18, b19, b20, b21}, our work represents a significant technical advancement in several key aspects. First, we uniquely integrate gene-gene interactions with gene expression profiles within the context of cell embedding, creating a dual-perspective representation that captures both direct transcriptional states and regulatory relationships. Second, we introduce a novel computational framework that simultaneously leverages data-driven gene-gene interactions and expression profiles to calculate cell embeddings, allowing for more nuanced detection of cell states. This integration provides distinct technical advantages: (1) enhanced ability to identify rare cell populations through preservation of subtle regulatory patterns, (2) improved robustness to technical noise by leveraging complementary data types, and (3) more biologically meaningful embeddings that reflect both expression levels and regulatory relationships. Our approach addresses a critical gap in current methods, which either focus solely on expression data or treat gene-gene interactions as separate entities, missing the complex interplay between these biological features. The novelty of our work lies not just in combining these data types, but in developing a computational framework that preserves and utilizes the unique information contained in both expression patterns and regulatory networks to create more comprehensive and accurate cell state representations. 

Unlike previous approaches that focus solely on gene expression, this method captures the regulatory relationships between genes, providing a more comprehensive embedding for downstream analysis, such as cell population discovery and rare cell type identification.

\section{Method}
\label{sec:sample2}

\begin{figure*}[htbp]
    \hspace{-1.7cm} % Add 1cm of horizontal space to the left
    \centering
    \includegraphics[width=0.9\linewidth] {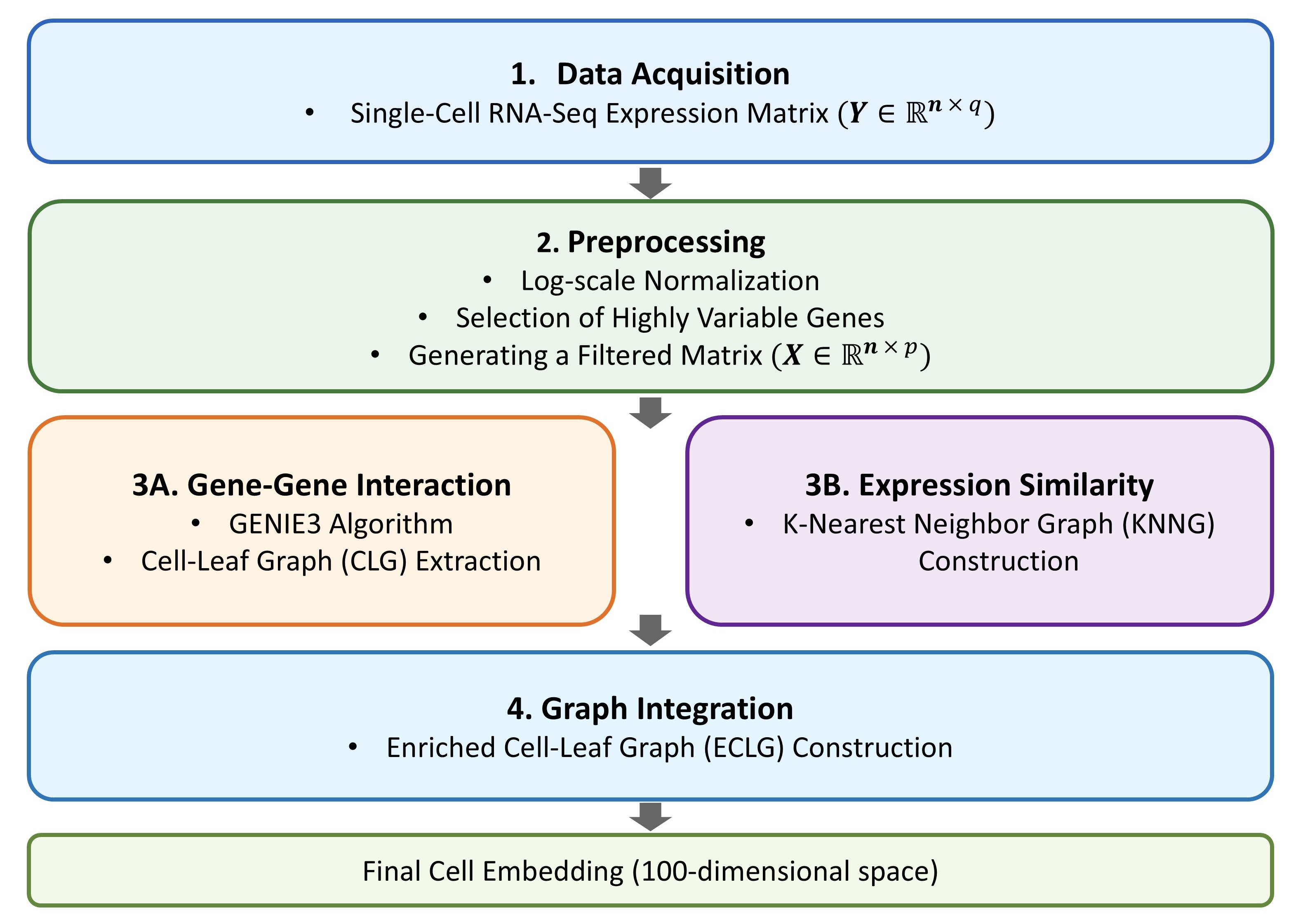}
    \hspace{-1.5cm} % Add 1cm of horizontal space to the right
    \caption{Enhanced Single-Cell RNA-Seq Embedding Workflow.}
    \label{fig1}
\end{figure*}

In this study, we propose a novel single-cell embedding method that integrates gene expression profiles with gene-gene interaction data to comprehensively represent cellular states. The process is depicted in ``Fig.~\ref{fig1}'', which outlines the workflow from data acquisition to the final cell embeddings used for downstream analysis. The workflow begins with the acquisition of gene expression profiles, followed by preprocessing steps such as log-scale normalization and filtering of highly variable genes. Subsequently, gene-gene interaction data is derived from the processed expression data. Let's first denote the initial gene expression matrix as \( Y \in \mathbb{R}^{n \times q} \), containing raw expression values for \( n \) cells across \( q \) genes. We apply log-scale normalization first, followed by selecting highly variable genes through filtering. Specifically, we retain the top 2000 genes based on variance, resulting in our working matrix \( X \in \mathbb{R}^{n \times p} \), where \( p < q \) represents the number of retained genes.

Following this, a \textbf{Cell-Leaf Graph (CLG)} \cite{b22} is constructed using the \textbf{GENIE3} algorithm to capture gene-gene interactions. In parallel, a \textbf{K-Nearest Neighbor Graph (KNNG)} \cite{b23} is generated based on gene expression similarities between cells. These two graphs are then integrated to form an \textbf{Enriched Cell-Leaf Graph (ECLG)}, which combines information from both gene expression profiles and gene-gene interactions.

A \textbf{Graph Neural Network} is applied to the ECLG to compute cell embeddings, preserving both gene interaction proximities and expression similarities. These embeddings are subsequently used for downstream analyses such as clustering, visualization, and trajectory inference. The following sections will provide a detailed explanation of each step in the workflow.

%% Figure 1

\subsection{Cell-Leaf Graph (CLG) extraction }
\label{sec:sample2.1}
\begin{figure*}[htbp]
    \vspace{-1.6cm} % Adjust the value to shift the figure up
    \hspace{-3.4 cm}
    \includegraphics[width=1.5\linewidth, height=1.23\linewidth] {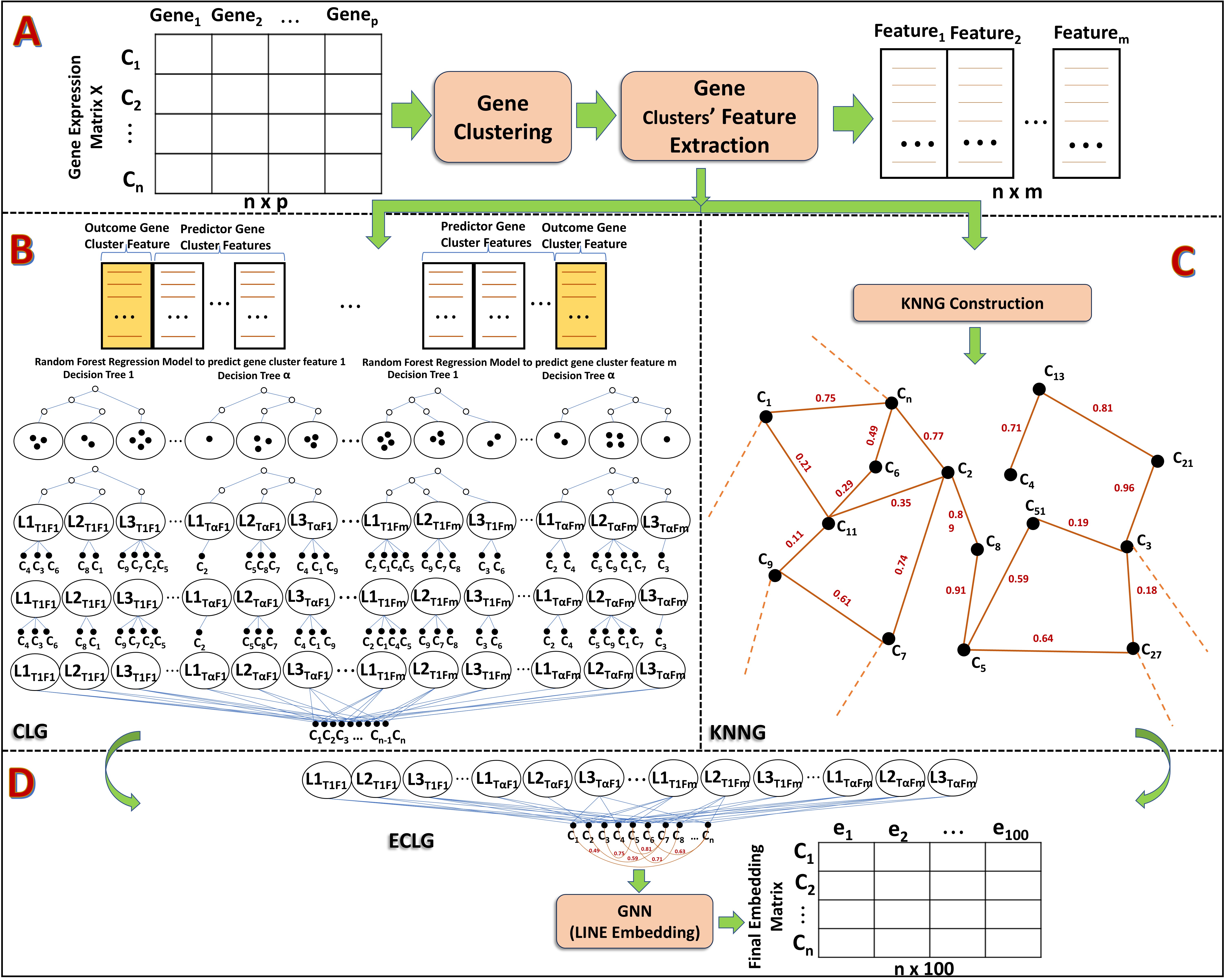}
    \caption{Overview of DAE. DAE takes a gene-expression matrix as its input and learns a non-linear representation of the cells. This learned embedding can be utilized for various downstream tasks. (A) Gene clusters’ signature (feature) extraction. (B) The extraction of a Cell-Leaf Graph is performed to capture the proximity of cells based on their interactions within co-expressed gene clusters. This graph modeling approach allows for the representation of cell proximities in terms of these interactions. (C) The K-Nearest Neighbor Graph (KNNG) is a widely employed data structure for depicting connections between individual cells, relying on their gene expression profiles. (D) In the end, both graphs are combined and a node embedding algorithm is applied to the fused graph in order to compute an embedding vector for each cell.}
    \label{fig2}
\end{figure*}

The first phase of our proposed embedding method is based on the GENIE3 algorithm \cite{b17} ``Fig.~\ref{fig2}'' part ``A''. Briefly, in this phase, each random forest is trained to predict the expression level of a target gene $x_i$ based on the expression levels of all other genes ($x_1,x_2,…,x_{(i-1)},x_{(i+1)},…  x_p)^T$,  in the dataset. The tree is constructed recursively by splitting the data into smaller subsets based on the expression levels of a randomly selected subset of genes. The splitting process continues until a stopping criterion is met, such as reaching a maximum depth or a minimum number of cells in each leaf. In our study, we set the minimum cell count in each leaf to 10 to ensure a greater likelihood of accurately representing rare cell populations.

In our study, gene-gene interactions are determined by considering all cells collectively using the GENIE3 algorithm. Please note that, in GENIE3, after training an ensemble of decision trees, candidate regulators of a target gene are ranked based on their frequency of appearance in the decision trees. Specifically, the importance score of a gene is calculated as the sum of the decrease in node impurity caused by the gene across all decision trees in which it appears as a splitting variable. As a result, a directed graph is created, with nodes denoting genes and edges denoting regulatory interactions. Gene A controls gene B, according to an edge from node A to node B. Although gene-gene interactions are inferred from the same gene expression data used for generating expression profiles, they represent distinct biological insights. These interactions, derived via the GENIE3 algorithm, capture the regulatory relationships between genes, providing a new layer of information that complements the raw gene expression data. In essence, this regulatory network reflects how genes influence each other’s expression, adding depth to the analysis that single-cell embeddings based on expression profiles alone do not capture.  

However, the final feature importance ranking and regulatory network extraction is not performed in this study. Instead, in the final layer of decision trees, we primarily concentrate on the relationships between the leaf nodes and the cells positioned beneath them. As is common knowledge, after training a random forest, samples (in our situation, cells) would be landed in various leaf nodes. Hence, two cells $C_i$ and $C_j$ should be placed in common or closer leaf nodes throughout our model if their gene-gene interactions are generally quite comparable. Hence, in this phase, we extract a relation graph by connecting cells to their relevant leaf nodes across all forests and then get rid of the rest of the trees. By doing so, we can model how each regulatory prediction model (i.e., each random forest regression model) relates to the others as well as how the decision trees inside each random forest interact with one another. The resulting graph is called Cell-Leaf Graph (CLG). The edge weights in the Cell-Leaf Graph (CLG) are assigned a value of 1, symbolizing each cell's strong connection to its corresponding leaf. We should mention that CLG is a heterogeneous bipartite graph where its nodes can be divided into two disjoint sets, U (cells) and V (Leaves) \cite{b24}. To put it concisely, in a heterogeneous bipartite graph, nodes within each part can symbolize diverse kinds of entities, causing nodes within the same part to be heterogeneous, or varied, such as leaf nodes originating from different decision trees.

%% Figure 2

\subsection{KNNG extraction}
\label{sec:sample2.2}

 A K-Nearest Neighbor Graph (KNNG) is a commonly used data structure to represent relationships among individual cells based on their gene expression profiles \cite{b25}. Each cell in the dataset is a node in the graph, and edges are drawn between each node and its K most similar nodes (i.e., the K "nearest neighbors" in terms of gene expression).
To create KNNG, as recommended in \cite{b25}, first, we apply PCA to convert the original data into vectors in a lower-dimensional space (50 PCs), then, for each cell, the K cells that have the smallest distances to it are selected as its nearest neighbors. For the K-Nearest Neighbor Graph (KNNG), we use the Gaussian Radial Basis Function (RBF) kernel to transform the distances between cells into similarity scores\cite{b26}. Given the distance \(d(u,v)\) between two cells \(u\) and \(v\), the RBF kernel computes a similarity score \(w_{\text{KNNG}}(u,v)\) as follows:
\begin{itemize}
    \item \textbf{Distance to Similarity Transformation:} \\
    The edge weights in the KNNG are based on the RBF kernel:

    \begin{equation}
    w_{\text{KNNG}}(u,v) = \exp\left(-\frac{d(u,v)^2}{2\sigma^2}\right)
    \end{equation}

    where \(d(u,v)\) is the Euclidean distance (or another distance metric) between the gene expression profiles of cells \(u\) and \(v\), and \(\sigma\) is the bandwidth parameter that controls the smoothness of the kernel. In our implementation, we set \(\sigma = 0.3\) to optimize the trade-off between local and global structure preservation. We determined this value by minimizing the Nearest Neighbor Error (NNE) across our experimental datasets.
    
    \item \textbf{Interpretation of Weights:} \\
    For small distances \(d(u,v)\), the similarity \(w_{\text{KNNG}}(u,v)\) approaches 1, indicating a strong connection between cells with similar gene expression profiles. For large distances \(d(u,v)\), the similarity decreases exponentially, approaching 0, reflecting weaker relationships between cells with dissimilar profiles.
\end{itemize}

In a KNNG graph, the neighbors of a cell are those cells with the most similar gene expression profiles, which often implies they are of the same or similar cell types or states. Therefore, a KNNG graph can capture the structure of the data and reveal the underlying cell populations in the dataset.

\subsection{Node Embedding on Integrated Graphs}
\label{sec:sample2.3}
 The integration of both the Cell-Leaf Graph (CLG) and K-Nearest Neighbor Graph (KNNG) is quite straightforward, given that the cells are shared in both graphs. Essentially, we augment the connectivity of the CLG by incorporating all edges from the KNNG, as the node set of KNNG falls under the subset of the CLG. Initially, the CLG exists as a heterogenous bipartite graph, but its merging with the KNNG — i.e., adding the KNNG links to it — transforms it into a general graph. This is due to the fact that certain cells within the same partition (U) become interconnected. The graph that emerges from this process is termed Enriched Cell-Leaf Graph (ECLG).
Next, a graph node embedding, also known as graph embedding or network representation learning \cite{b27} is applied to ECLG. Generally, graph node embedding is a technique for learning low-dimensional vector representations, or embeddings, of nodes in a graph. The goal of graph node embedding is to preserve the structural and semantic properties of the graph, such as node proximity, connectivity patterns, and node attributes, in the learned embeddings. In this work, we use the LINE algorithm \cite{b28}. We opt to utilize LINE because it's deliberately designed to efficiently handle huge networks, a vital requirement for a multitude of real-world applications. LINE optimizes an objective function that preserves both first-order and second-order proximities in the embedding space:

\begin{itemize}
    \item \textbf{First-order proximity:} \\
    The first-order proximity represents the direct connections between nodes. For nodes \(u\) and \(v\) connected by an edge, LINE minimizes the following loss function:
    %\[
    %L_1 = - \sum_{(u,v) \in E} w(u,v) \log \sigma(z_u^T z_v)
    %\]
    \begin{equation}
        L_1 = - \sum_{(u,v) \in E} w(u,v) \log \sigma(z_u^T z_v)
    \end{equation}
    
    where \(\sigma(x) = \frac{1}{1 + e^{-x}}\) is the sigmoid function, and \(z_u\) and \(z_v\) are the embedding vectors for nodes \(u\) and \(v\), respectively.

    \item \textbf{Second-order proximity:} \\
    The second-order proximity captures the neighborhood similarity between nodes. For two nodes that share many common neighbors, the second-order proximity ensures their embeddings are similar, even if they aren't directly connected. The objective is to minimize:
    
    %\[
    %L_2 = - \sum_{u \in V} \sum_{v \in N(u)} p(v \mid u) \log \sigma(z_u^T z_v)
    %\]
    \begin{equation}
        L_2 = - \sum_{u \in V} \sum_{v \in N(u)} p(v \mid u) \log \sigma(z_u^T z_v)
    \end{equation}
    
    where \(N(u)\) represents the neighbors of node \(u\), and \(p(v \mid u)\) is the conditional probability of reaching node \(v\) from node \(u\).
\end{itemize}
By optimizing both \(L_1\) and \(L_2\), LINE preserves both local and global structures in the embedding space, producing a low-dimensional representation of each cell that reflects both its gene expression similarity and its gene-gene interaction information. For a more in-depth understanding, consider referring to the original research paper \cite{b28}.

The ultimate cell embedding serves as a fresh latent representation of our cells. This final embedding encompasses not only the proximity of cells’ gene expression profiles but also takes into account the similarity between cells in terms of their gene-gene interaction patterns across the data. Finally, these resulting embeddings can be used for any downstream single-cell analysis like visualization, clustering, and trajectory detection.

\subsection{Extension of Dual Aspect Embedding for large-scale datasets}
\label{sec:sample2.4}
It is now possible to profile tens of thousands of individual cells in a single massively parallel experiment because of recent developments in scRNA-seq technology. Despite the potentially time-intensive nature of extracting the Cell-Leaf Graph (CLG) using Random Forests, it's worth noting that this process can be highly parallelizable due to the inherent structure of random forests. In summary, within a parallel computing environment, individual cores can be allocated the task of creating one or more trees. By executing these tasks simultaneously, the overall process is greatly expedited compared to carrying them out sequentially. Nonetheless, the focus of this part is to introduce a modified version of proposed algorithm that has been specifically optimized for efficient execution on standard personal computers. To do so, instead of using whole gene expression matrix \(X\) as input, we extract features from subspaces of \(X\), as previously examined in \cite{b13}, as follows: 

 Initially, the expression matrix is transposed, treating genes as observations and cells as features, and then subjected to Principal Component Analysis (PCA). The 'elbow method' is employed to determine and retain the principal components that offer the most informative insights \cite{b35}. Following this, k-means++ clustering \cite{b36} is used on this condensed gene representation, resulting in gene clusters: \{$C_1, C_2 ,,, C_k$\}. In this clustering, we ascertain the number of clusters by identifying the 'elbow point' on a graph that plots the sum of squared errors against an increasing number of clusters. Subsequently, we execute PCA again on every gene cluster $C_i$ (with cells presented as rows and genes grouped in that cluster as columns), retaining the most significant principal components using the elbow technique. This procedure results in K feature matrices, where each cell is characterized by its principal components derived from K gene clusters. We then create a final feature matrix \(F\) by concatenating matrices derived from individual gene clusters:\begin{equation}
F=[F_1,F_2,…,F_K]
\end{equation}

 We assume the final feature matrix \(F\) has \(K\) distinct features, where \(K\) represents the number of gene clusters. Now, we can use \(F\) as the input to proposed method. While it is noteworthy to state that the final feature matrix \(F\) does not contain actual gene expression values, it does have features extracted from gene clusters (genes are grouped into clusters based on shared expression patterns) and we could state that these features are unique signatures derived from clusters of genes. Therefore, by building a Cell-Leaf Graph (CLG) from these clusters’ features, we are representing the interactions that exist between the gene clusters. This method can not only streamline our analysis but also possibly enhance its resilience to the intrinsic noise found in single-cell data (dropouts). Also, If the gene clusters reflect meaningful biological categories, such as co-regulated genes or genes involved in the same pathway, this approach could efficiently capture the major differences between cell types in downstream analyses like clustering and visualization.
 In addition, the same features can be used to create KNNG to model relationships among individual cells based on their gene expression profiles. Thus, we could argue that the entire process is essentially an approximation of the original embedding approach. We call our proposed method:  Dual Aspect Embedding (DAE). ``Fig.~\ref{fig2}'' provides a visual representation of the step-by-step process of DAE.

\subsection{Technical Considerations}
\label{sec:sample2.5}
The stability of the constructed gene network, despite the inherent randomness in the random forest algorithm, is ensured through several strategies. We use an ensemble of decision trees, which mitigates the impact of randomness by aggregating results from multiple trees. We focus on relationships between leaf nodes and cells, ensuring more consistent pattern capture across the ensemble. A minimum cell count criterion for each leaf ensures an accurate representation of rare cell populations. Additionally, running the model multiple times with different random seeds and averaging the results further stabilizes the network, reducing the impact of random variations.

While variation in gene networks can have some impact on the final results, our methodology mitigates this through several mechanisms. By integrating both gene expression profiles and gene-gene interaction information, we ensure that key biological interactions are consistently captured. Our embedding method combines the Cell-Leaf Graph (CLG) and the K-Nearest Neighbor Graph (KNNG), enhancing robustness by considering both gene interaction proximities and expression similarities. For example, when varying the number of trees in the random forest (from 50 to 500), the Nearest Neighbor Error (NNE) values remained stable, reaching a plateau with more than 200 trees per forest, as shown in experimental result (section 3.4). Similarly, running the model with different random seeds resulted in an average NNE fluctuation of only 0.8\% across datasets, demonstrating the robustness of the method against random variations.

Please note that the proposed approach aims to extract a novel embedding that integrates gene expression profiles and gene-gene interactions, rather than merely performing dimensionality reduction for summarization or visualization. By combining these two data types, we create a comprehensive representation of cellular states, capturing the complex interplay between genes. This enriched embedding space enhances the ability to distinguish between different cell types and states, providing a robust foundation for various downstream single-cell analyses such as clustering, visualization, and trajectory inference. This approach addresses the limitations of traditional methods that treat gene expression data in isolation, offering a more biologically meaningful and holistic view of cellular heterogeneity and dynamics.

To implement DAE, we employed the gene clusters’ feature construction module presented in \cite{b13} for gene clusters' feature engineering, the "RandomForest" R package \cite{b42} for the construction of CLG, while the KNNG construction leverages the R FNN package \cite{b43}. Additionally, for the node embedding, we made use of the rline package, which can be found at:
\\
https://github.com/YosefLab/Rline.
 
\section{Experimental Results}
\label{sec:sample3}
In this study, we utilized six distinct datasets as our benchmarking standards (``Table.~\ref{Table1}''). These datasets were chosen based on a few important factors, including their high label confidence and diversity of cells in terms of developmental phases and environmental variables. We then utilize them as our "gold standard" for comparison. In summary, all datasets underwent normalization and log transformation. Subsequently, we identified the top 2000 genes with high variability as a preliminary step before initiating any data analysis.
Single-cell embedding techniques can be broadly classified into two categories: (i) Direct transformation methods, such as PCA, SVD, Kernel PCA (kPCA), t-SNE, and scVI \cite{b9} that convert high-dimensional single-cell data directly into a lower-dimensional space; (ii) Similarity-based methods like RAFSIL \cite{b13} and SIMLR \cite{b8}, which first compute a similarity or dissimilarity matrix from the data, and subsequently use this matrix as an input to techniques like t-SNE to extract a lower-dimensional representation of the data. DAE, falling into the first category, had its performance benchmarked against techniques from both groups in our study. However, for visualization purposes, this DAE embedding is subsequently inputted into t-SNE for conversion into a two-dimensional space. In our research, we empirically set the DAE's embedding dimensions to 100. We distinguish between three different scenarios: similarity learning, dimension reduction for visualization, and clustering. All of these have important roles to play in the analysis, visualization, and interpretation of scRNA-seq data, but they also have different goals, therefore we evaluate them separately. 

%% Table 1
\begin{table*}[htbp]
    \centering
    \caption{Datasets}
    \begin{tabular}{|c|c|c|c|c|c|c|}
    \hline
    \multirow{2}{*}{Datasets} & \multirow{2}{*}{\# cells} & \multirow{2}{*}{\# genes} & \multirow{2}{*}{\# cell type} & \multirow{2}{*}{Sparsity (in \%)} & \multirow{2}{*}{Unit} & \multirow{2}{*}{Reference} \\
     & & & & & & \\ \hline
    Usoskin &622	&25334	&4	&85	&TPM	&\cite{b29}\\ \hline
    Cortex &3005	&19972	&9	&81	&UMI	&\cite{b30}\\ \hline
    Macoscko &10559	&23288	&39	&90	&UMI	&\cite{b31}\\ \hline
    Chen &14437	&23284	&47	&93	&UMI	&\cite{b32}\\ \hline
    Campbel &20921	&26774	&20	&93	&UMI	&\cite{b33}\\ \hline
    PBMC &76899	&32738	&7	&98	&UMI	&\cite{b34}\\ \hline
    \end{tabular}
    \label{Table1}
\end{table*}

\subsection{Similarity Learning}
\label{sec:sample3.1}
In this analysis, to evaluate the efficiency of the obtained embedding, we measure the Nearest Neighbor Error (NNE). NNE is defined as the proportion of cells whose nearest neighbors (in the embedding space) are of a different cell type. Mathematically, NNE is calculated as follows:

\begin{equation}
\text{NNE} = \frac{1}{N} \sum_{i=1}^{N} \mathbb{I}(y_i \neq \hat{y}_i)
\end{equation}

where \( N \) is the total number of cells, \( y_i \) is the true cell type of the \( i \)-th cell, \( \hat{y}_i \) is the cell type of the nearest neighbor of the \( i \)-th cell in the embedding space, and \( \mathbb{I} \) is the indicator function that returns 1 if the statement is true and 0 otherwise. A lower NNE indicates that the embedding better preserves the local structure of the data, as cells of the same type are more likely to be close to each other.

 We juxtapose our suggested approach with three techniques - RAFSIL, scVI, and SIMLR - expressly tailored for scRNA-seq data. This is in addition to contrasting DAE with a range of conventional similarity or dissimilarity measures computed by Euclidean, Spearman, and Pearson correlation, along with popular dimension reduction methods such as t-Distributed Stochastic Neighbor Embedding (t-SNE), Principal Component Analysis (PCA), Singular Value Decomposition (SVD), and Kernel PCA (kPCA). For the latter three methods, throughout this research, we have conformed to previous recommendations by reducing the dimensionality to 50 \cite{b25}. ``Table.~\ref{Table2}'' provides a summary of our findings. We observe that DAE and RAFSIL acquire similarities that better describe annotated cell populations (i.e., lower NNE).
We also find that SIMLR's performance decreases with increasing data size.

 %% Table 2
\begin{table*}[htbp]
    \centering
    \caption{Nearest neighbor error values for similarity learning (in percent, lower is better)}
    \begin{tabular}{|c|c|c|c|c|c|c|}
    \hline
    \multirow{2}{*}{Method} & \multirow{2}{*}{Usoskin} & \multirow{2}{*}{Cortex} & \multirow{2}{*}{Macoscko} & \multirow{2}{*}{Chen} & \multirow{2}{*}{Campbel} & \multirow{2}{*}{PBMC} \\
     & & & & & & \\ \hline
    DAE &\textbf{1.4}	&\textbf{3.5}	&\textbf{4.2}	&\textbf{7.1}	&12.2	&\textbf{8.5}\\ \hline
    RAFSIL &2.6	&4.2	&5.1	&9.4	&11.3	&12.3\\ \hline
    scVI &3.2	&4.2	&4.5	&11.2	&11.3	&10.1\\ \hline
    SIMLR &2.9	&3.9	&17.7	&19.5	&26.2	&27.1\\ \hline
    PCA &5.2	&16.2	&20.3	&21.6	&23.5	&27.1\\ \hline
    SVD &5.1	&14.7	&18.5	&23.3	&18.2	&17.7\\ \hline
    kPCA &8.5	&23.6	&24.4	&27.6	&17.2	&27.5\\ \hline
    Euclidean &4.1	&13.5	&22.1	&27.6	&32.4	&26.5\\ \hline
    Pearson &4.7	&13.5	&17.5	&22.4	&29.5	&24.9\\ \hline
    Spearman &7.4	&14.7	&16.5	&25.8	&25.7	&22.4\\ \hline
    \end{tabular}
    \label{Table2}
\end{table*}

%% Table 3
\begin{table*}[htbp]
    \centering
    \caption{Nearest neighbor error values for visualization with t-SNE (in percent, lower is better)}
    \begin{tabular}{|c|c|c|c|c|c|c|}
    \hline
    \multirow{2}{*}{Method} & \multirow{2}{*}{Usoskin} & \multirow{2}{*}{Cortex} & \multirow{2}{*}{Macoscko} & \multirow{2}{*}{Chen} & \multirow{2}{*}{Campbel} & \multirow{2}{*}{PBMC} \\
     & & & & & & \\ \hline
    DAE &\textbf{0.77}	&3.3	&\textbf{2.8}	&\textbf{3.4}	&7.4	&\textbf{5.9}\\ \hline
    RAFSIL &0.93	&5.7	&\textbf{2.8}	&6.2	&9.2	&9.2\\ \hline
    scVI &1.2	&5.3	&3.2	&7.5	&\textbf{5.8}	&8.4\\ \hline
    SIMLR &2.3	&\textbf{3.1}	&13.1	&16.3	&22.2	&25.3\\ \hline
    PCA &4.2	&7.1	&7.7	&15.3	&9.4	&20.5\\ \hline
    SVD &4.7	&7.6	&7.3	&14.2	&10.3	&12.6\\ \hline
    kPCA &6.3	&6.3	&8.5	&15.2	&13.5	&23.4\\ \hline
    Euclidean &5.1	&6.2	&7.8	&13.8	&8.8	&17.7\\ \hline
    Pearson &4.6	&6.8	&7.3	&13.1	&11.3	&15.1\\ \hline     
    Spearman &4.1	&6.3	&7.3	&12.5	&9.7	&16.2\\ \hline
    Monocle &5.3    &6.2    &7.9    &11.3   &8.3    &13.7\\ \hline
    Seurat (UMAP) &5.5    &6.2    &7.9    &11.1   &8.5    &13.4\\ \hline
    \end{tabular}
    \label{Table3}
\end{table*}

\subsection{Dimension Reduction for Visualization}
\label{sec:sample3.2}
To reduce dimensions, we utilized t-SNE on the acquired embedding and/or dissimilarity matrix generated by several methods such as DAE, SIMLR, scVI, RAFSIL, PCA, SVD, kPCA, spearman, Pearson, and Euclidean. In addition, we utilize Seurat \cite{b37} and Monocle \cite{b38}, two widely recognized tools for single-cell analysis, to incorporate all datasets for a more extensive examination. We used UMAP for Seurat instead of t-SNE due to its superior ability to preserve both local and global data structures, faster computation speed, and better scalability for large datasets. UMAP also produces clearer visualizations and better-defined clusters, which are crucial for accurately identifying cell populations in single-cell RNA-seq data. Additionally, UMAP is widely adopted and recommended in the single-cell analysis community, ensuring that our results are in line with current best practices.

In this experiment, the NNE served as a quality metric, utilizing Euclidean distances in the reduced-dimensional space for all methods. The results are summarized in ``Table.~\ref{Table3}''. The results indicate that the DAE yields an average NNE of 4.09\%, outperforming both domain-specific approaches and traditional embedding techniques. Additionally, as indicated in ``Table.~\ref{Table3}'', Seurat and Monocle yield comparable NNEs, given that UMAP is a fundamental visualization technique employed in both methods.

\begin{figure*}[htbp]
  \centering
  
  \begin{minipage}[b]{0.24\textwidth}
    \centering
    \includegraphics[width=\linewidth]{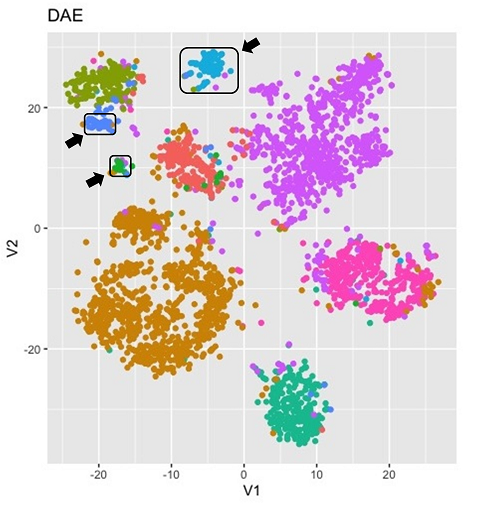}
    % \subcaption{Subfigure 1}
    \label{fig:subfig1}
  \end{minipage}%
  \hfill
  \begin{minipage}[b]{0.24\textwidth}
    \centering
    \includegraphics[width=\linewidth]{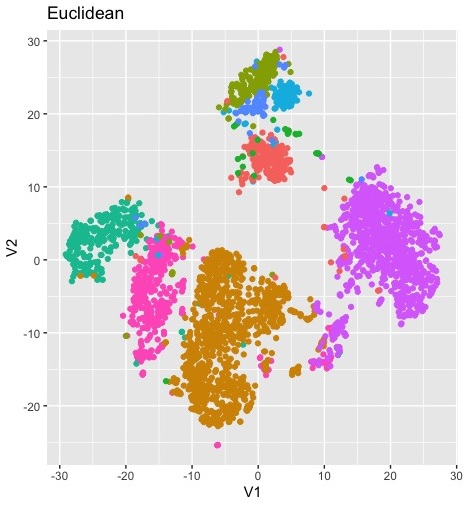}
    % \subcaption{Subfigure 2}
    \label{fig:subfig2}
  \end{minipage}%
  \hfill
  \begin{minipage}[b]{0.24\textwidth}
    \centering
    \includegraphics[width=\linewidth]{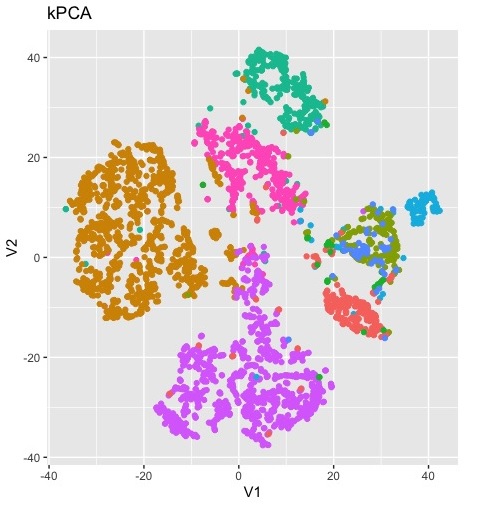}
    % \subcaption{Subfigure 3}
    \label{fig:subfig3}
  \end{minipage}%
  \hfill
  \begin{minipage}[b]{0.24\textwidth}
    \centering
    \includegraphics[width=\linewidth]{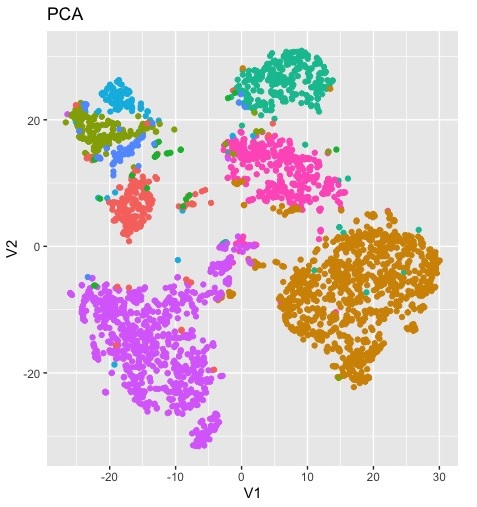}
    % \subcaption{Subfigure 4}
    \label{fig:subfig4}
  \end{minipage}
  
  \vspace{0.5cm}
  
  \begin{minipage}[b]{0.24\textwidth}
    \centering
    \includegraphics[width=\linewidth]{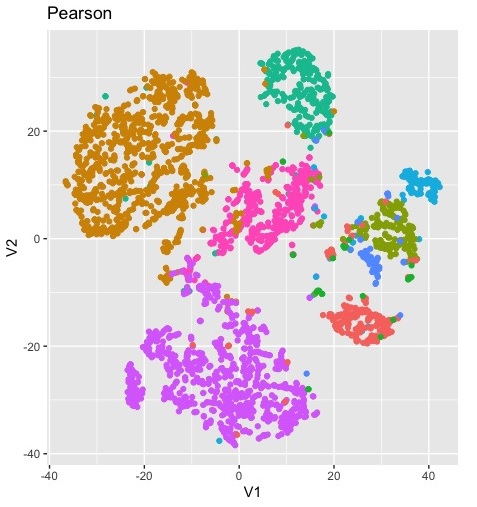}
    % \subcaption{Subfigure 5}
    \label{fig:subfig5}
  \end{minipage}%
  \hfill
  \begin{minipage}[b]{0.24\textwidth}
    \centering
    \includegraphics[width=\linewidth]{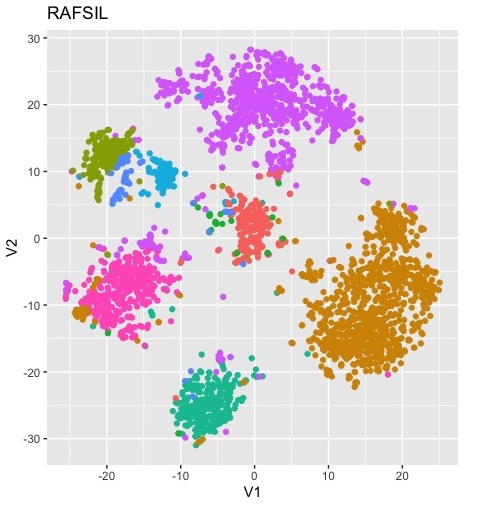}
    % \subcaption{Subfigure 6}
    \label{fig:subfig6}
  \end{minipage}%
  \hfill
  \begin{minipage}[b]{0.24\textwidth}
    \centering
    \includegraphics[width=\linewidth]{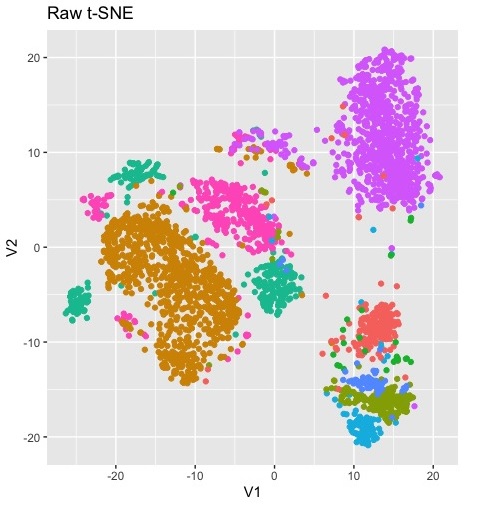}
    % \subcaption{Subfigure 7}
    \label{fig:subfig7}
  \end{minipage}%
  \hfill
  \begin{minipage}[b]{0.24\textwidth}
    \centering
    \includegraphics[width=\linewidth]{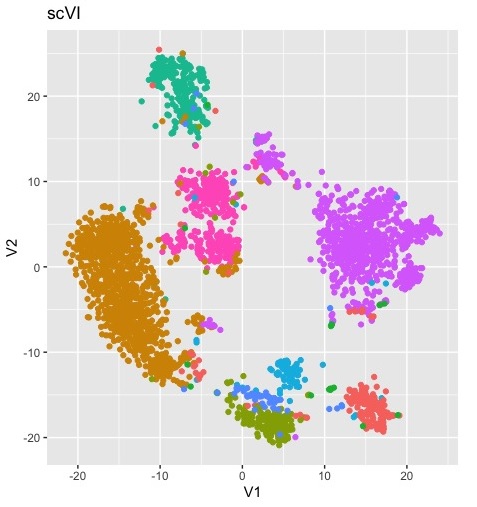}
    % \subcaption{Subfigure 8}
    \label{fig:subfig8}
  \end{minipage}
  
  \vspace{0.5cm}
  
  \begin{minipage}[b]{0.24\textwidth}
    \centering
    \includegraphics[width=\linewidth]{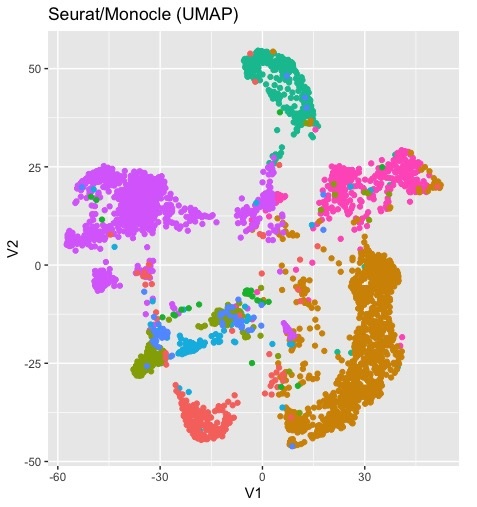}
    % \subcaption{Subfigure 9}
    \label{fig:subfig9}
  \end{minipage}%
  \hfill
  \begin{minipage}[b]{0.24\textwidth}
    \centering
    \includegraphics[width=\linewidth]{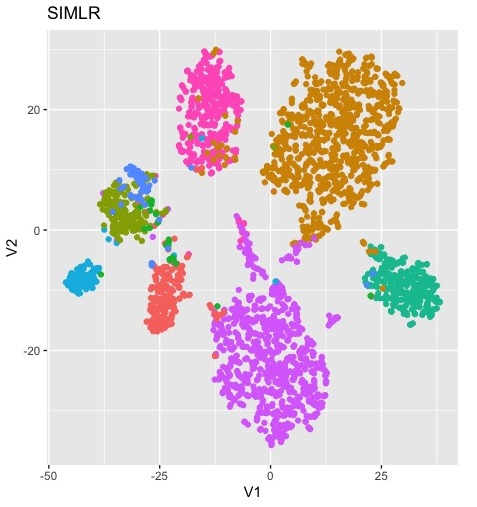}
    % \subcaption{Subfigure 10}
    \label{fig:subfig10}
  \end{minipage}%
  \hfill
  \begin{minipage}[b]{0.24\textwidth}
    \centering
    \includegraphics[width=\linewidth]{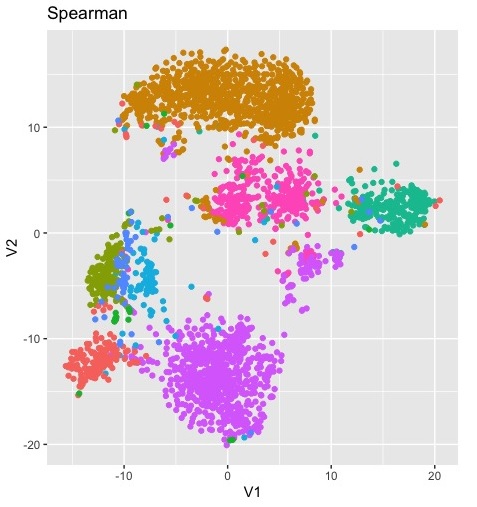}
    % \subcaption{Subfigure 11}
    \label{fig:subfig11}
  \end{minipage}%
  \hfill
  \begin{minipage}[b]{0.24\textwidth}
    \centering
    \includegraphics[width=\linewidth]{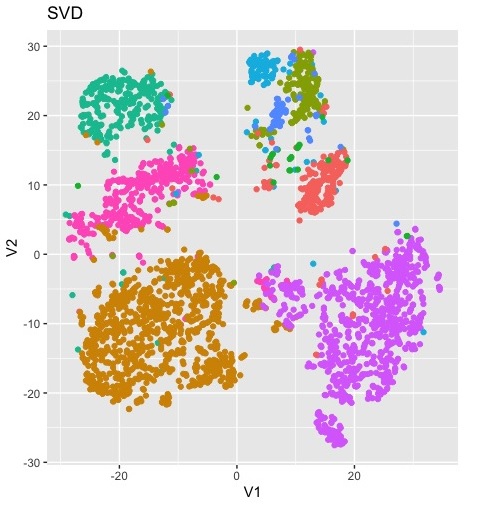}
    % \subcaption{Subfigure 12}
    \label{fig:subfig12}
  \end{minipage}
  
  \vspace{0.5cm}
  
  \begin{minipage}[b]{\textwidth}
    \centering
    \includegraphics[width=\linewidth]{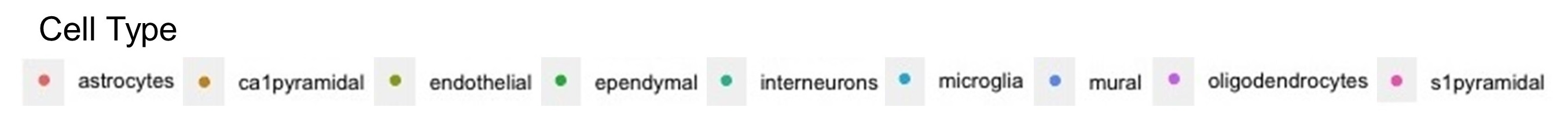}
    % \subcaption{Legend}
    \label{fig:legend}
  \end{minipage}

  \caption{The figure presents 2D visualization plots for various embedding methods, where each point represents a cell. The spatial arrangement of the points is determined using t-SNE/UMAP for the Cortex dataset. In the observed plots, rare cell types (mural, microglia, and ependymal) are visible across several methods. However, the DAE plot emphasizes their distinction more clearly by preserving their separation within the embedding space. These cell types are highlighted and delineated by bounding boxes within the DAE plot, which improves their visualization.}
  \label{fig3}
\end{figure*}

We have noted that DAE embedding enhances the likelihood of detecting rare cell types. To substantiate this assertion, we concentrate on the Cortex data, which was sourced from the mouse cortex and hippocampus, comprising 9 principal cell types: interneurons, s1pyramidal, ca1pyramidal, oligodendrocytes, microglia, endothelial, astrocytes, ependymal, and mural \cite{b30}. Within these, microglia (0.03\%), ependymal (0.008\%), and mural (0.02\%) represent the rare cell populations. We applied various embedding techniques, followed by t-SNE visualization (as depicted in ``Fig.~\ref{fig3}''). The proposed method integrates gene expression profiles with gene-gene interaction information, significantly enhancing the identification of rare cell types such as microglia, ependymal cells, and mural cells. Microglia, the resident immune cells of the CNS, ependymal cells involved in cerebrospinal fluid regulation, and mural cells associated with vascular stability, each present unique identification challenges due to their rare presence and subtle gene expression profiles. By capturing the intricate regulatory networks and gene-gene interactions specific to these cells, the method preserves critical biological signals during dimensionality reduction, improving clustering and visualization. This leads to a more accurate segregation of rare cell types from the majority, as demonstrated by the distinct bounding boxes in the DAE plot, thereby offering a robust framework for their detection and analysis in single-cell RNA-seq datasets.

%% Figure 3

\begin{figure}[htbp]
\centerline{\includegraphics[width=\linewidth]{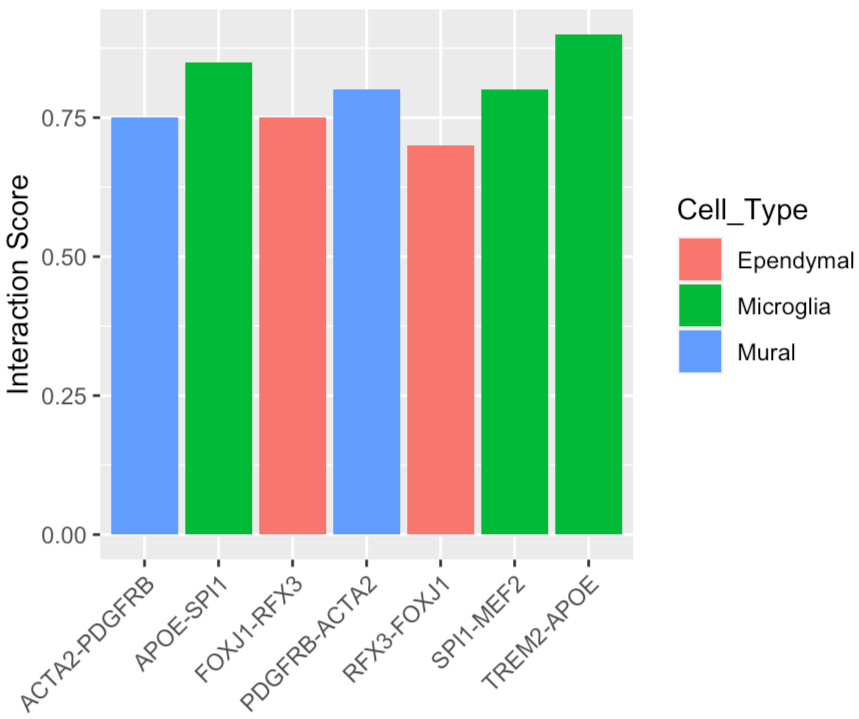}}
\caption{Significant gene-gene interaction scores for 3 rare cell types in mouse cortex and hippocampus }
\label{Fig4_gene-gene-int}
\end{figure}

Furthermore, a bar plot analysis (as shown in ``Fig.~\ref{Fig4_gene-gene-int}'') of gene-gene interaction scores extracted from CLG highlights the significant interactions for these rare cell types. For microglia, critical interactions involve genes such as TREM2 and APOE, which interact to shift microglia from a homeostatic to a neurodegenerative state, crucial in Alzheimer's disease. Additionally, interactions between APOE, SPI1, and MEF2 play vital roles in regulating microglia function during inflammation. For ependymal cells, interactions between FOXJ1 and RFX3 are essential for the development and function of ciliated ependymal cells involved in cerebrospinal fluid regulation. In mural cells, interactions between PDGFRB and ACTA2 are key for vascular stability and development. The bar plot demonstrates the strength of these interactions, with interaction scores reflecting the robustness of these regulatory networks. This detailed visualization underscores the method's capability to effectively capture and utilize gene-gene interactions, providing deeper insights into the identification and analysis of rare cell types in single-cell RNA-seq datasets.

%% Figure 4
 
The outcomes indicate that, in terms of offering a superior visualization for rare cell types within the resultant 2D map, DAE outperforms the others. However, our findings indicate that SIMLR produces a more compact depiction of the labeled subpopulations, surpassing the performance of the rest of the methods in this experiment. This outcome was anticipated because SIMLR is designed to explicitly generate a condensed representation of the data within a specified number of clusters.

\subsection{Clustering}
\label{sec:sample3.3}
\begin{figure*}[htbp]
  \centering
  
  \begin{subfigure}{0.49\linewidth}
    \centering
    \includegraphics[width=\linewidth]{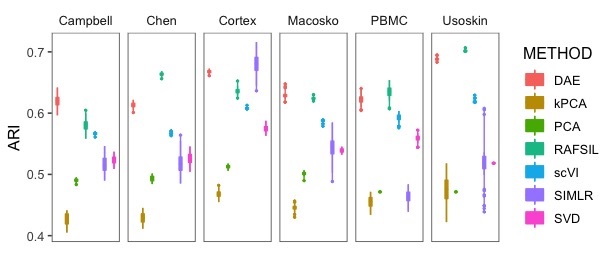}
    \caption{DBSCAN}
    \label{fig:subfig1}
  \end{subfigure}%
  \hfill
  \begin{subfigure}{0.49\linewidth}
    \centering
    \includegraphics[width=\linewidth]{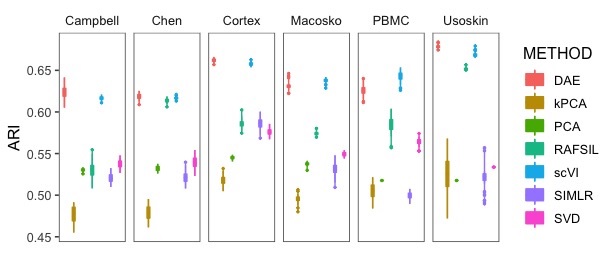}
    \caption{Kmeans++}
    \label{fig:subfig2}
  \end{subfigure}
  
  \vspace{0.5cm}
  
  \begin{subfigure}{0.49\linewidth}
    \centering
    \includegraphics[width=\linewidth]{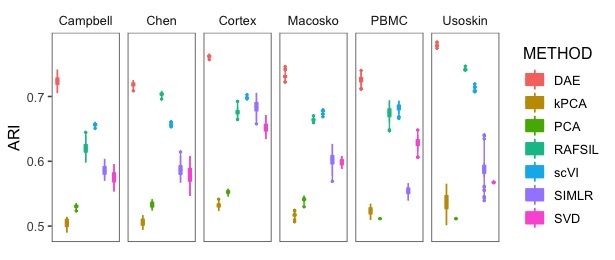}
    \caption{Phenograph}
    \label{fig:subfig3}
  \end{subfigure}%
  \hfill
  \begin{subfigure}{0.49\linewidth}
    \centering
    \includegraphics[width=\linewidth]{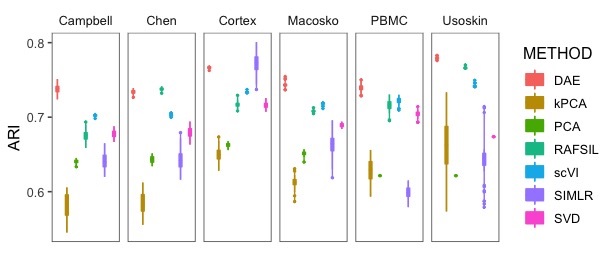}
    \caption{SC3}
    \label{fig:subfig4}
  \end{subfigure}
  
  \vspace{0.5cm}
  
  \begin{subfigure}{0.49\linewidth}
    \centering
    \includegraphics[width=\linewidth]{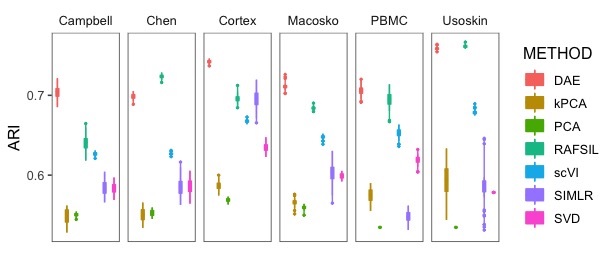}
    \caption{Seurat}
    \label{fig:subfig5}
  \end{subfigure}%
  \hfill
  \begin{subfigure}{0.49\linewidth}
    \centering
    \includegraphics[width=\linewidth]{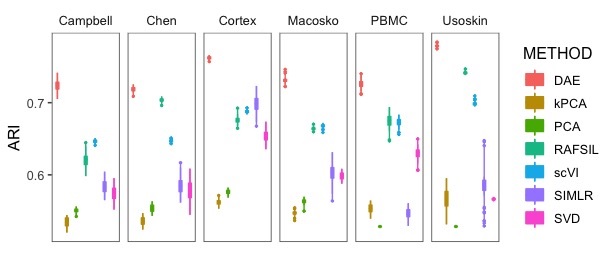}
    \caption{Monoloce}
    \label{fig:subfig6}
  \end{subfigure}
  \caption{DAE enhances the performance of single-cell clustering methods. Each sub-figure in the figure contains box plots that represent
the Adjusted Rand Index (ARI) of a specific clustering technique applied to six datasets}
  \label{fig5}
\end{figure*}

\begin{figure*}[t] % <-- Changed 'htbp' to 't'
  \centering
  
  \begin{subfigure}{0.49\linewidth}
    \centering
    \includegraphics[width=\linewidth]{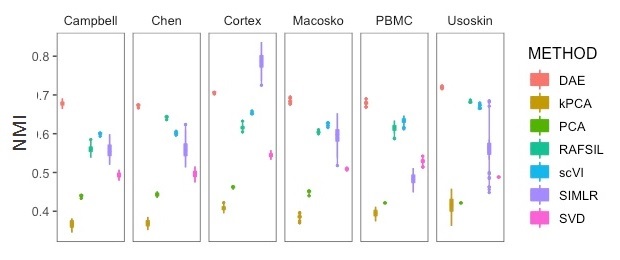}
    \caption{DBSCAN}
    \label{fig:subfig1}
  \end{subfigure}%
  \hfill
  \begin{subfigure}{0.49\linewidth}
    \centering
    \includegraphics[width=\linewidth]{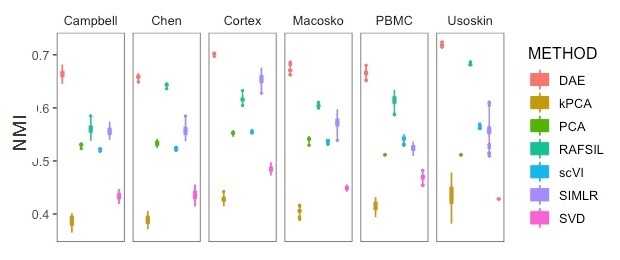}
    \caption{Kmeans++}
    \label{fig:subfig2}
  \end{subfigure}
  
  \vspace{0.5cm}
  
  \begin{subfigure}{0.49\linewidth}
    \centering
    \includegraphics[width=\linewidth]{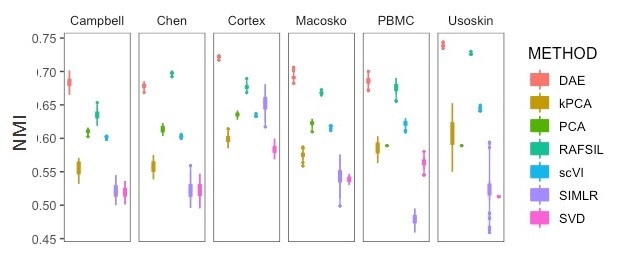}
    \caption{Phenograph}
    \label{fig:subfig3}
  \end{subfigure}%
  \hfill
  \begin{subfigure}{0.49\linewidth}
    \centering
    \includegraphics[width=\linewidth]{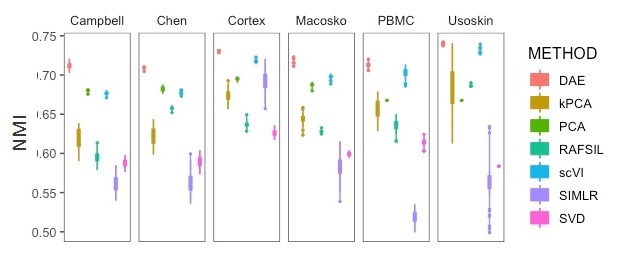}
    \caption{SC3}
    \label{fig:subfig4}
  \end{subfigure}
  
  \vspace{0.5cm}
  
  \begin{subfigure}{0.49\linewidth}
    \centering
    \includegraphics[width=\linewidth]{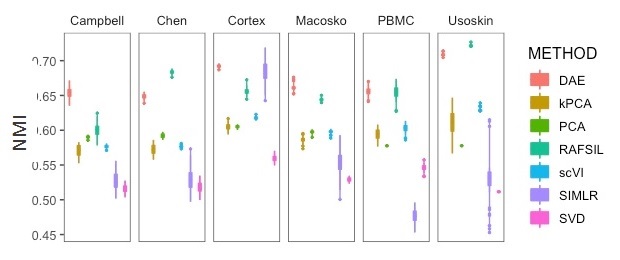}
    \caption{Seurat}
    \label{fig:subfig5}
  \end{subfigure}%
  \hfill
  \begin{subfigure}{0.49\linewidth}
    \centering
    \includegraphics[width=\linewidth]{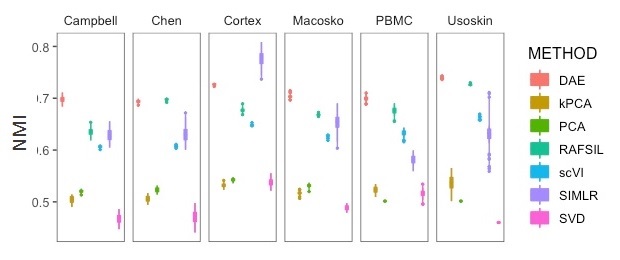}
    \caption{Monoloce}
    \label{fig:subfig6}
  \end{subfigure}
  \caption{DAE enhances the performance of single-cell clustering methods. Each sub-figure in the figure contains box plots that represent the Normalized Mutual Information (NMI) of a specific clustering technique applied to six datasets}
  \label{fig6}
\end{figure*}

Next, we delved into evaluating our embedding's performance relative to cell clustering, a key step in most scRNA-seq data analyses and a commonly used approach to uncover cell populations. We evaluated the performance of six unsupervised clustering algorithms: Kmeans++ \cite{b28}, Phenograph \cite{b25}, DBSCAN \cite{b39}, SC3 \cite{b40}, Seurat , and Monocle in identifying cell types throughout our benchmarks. Of these methods, Kmeans++ requires a predefined number of clusters. Therefore, we supplied the correct number of clusters in advance for Kmeans++.
Briefly, we run each embedding method (DAE, RAFSIL, SIMLIR, scVI, PCA, SVD, and kPCA) on datasets, followed by clustering. Please note that the quality of the clustering was then gauged by determining the Adjusted Rand Index (ARI) and Normalized Mutual Information (NMI)\cite{b41} between the assignments and true labels. Notably, certain methods, such as Kmeans++ and SC3, introduce randomness. To account for this and for assessing the stability of the clustering outcomes, we conducted a series of tests where we randomly removed 15\% of the cells from each dataset. We then performed each clustering method 30 times on these modified datasets, and the results were reported in ``Fig.~\ref{fig5}'' and ``Fig.~\ref{fig6}''. 
The outcomes indicate that DAE enhances the performance of clustering methods (on average higher ARI and NMI for most of the clustering methods). Furthermore, our findings demonstrate that DAE, RAFSIL, and scVI generally outperform traditional embeddings (PCA, SVD, kPCA). Furthermore, even though we set the precise number of clusters for Kmeans++ beforehand, its performance did not meet expectations for certain benchmarks. Kmeans++ assumes that clusters are convex, which means that it performs best with spherical clusters of similar size. It may not be robust for data that does not meet these assumptions, such as data with irregularly shaped cell clusters. Kindly note that both Seurat and Monocle utilize a graph community detection clustering method similar to Phenograph, albeit with slight modifications (such as using different versions of PCA and Louvain community detection). Consequently, the clustering results they produce are similar to each other. As can be seen in ``Fig.~\ref{fig5}'',  in our analysis, we also evaluated the level of fluctuation by computing the interquartile range (IQR) for each resampling iteration with each embedding method, and then we averaged this across all datasets, defining this measure as an IQR. Our results indicated that PCA offered the most consistent outcomes, with an average IQR of 2.74\%, while SIMLR exhibited the least stability, with a higher average IQR of 9.8\%.

%% Figure 5

%% Figure 6

\subsection{Sensitivity analysis for DAE}
\label{sec:sample3.4}
\begin{figure}[htbp]
\centerline{\includegraphics[width=\linewidth]{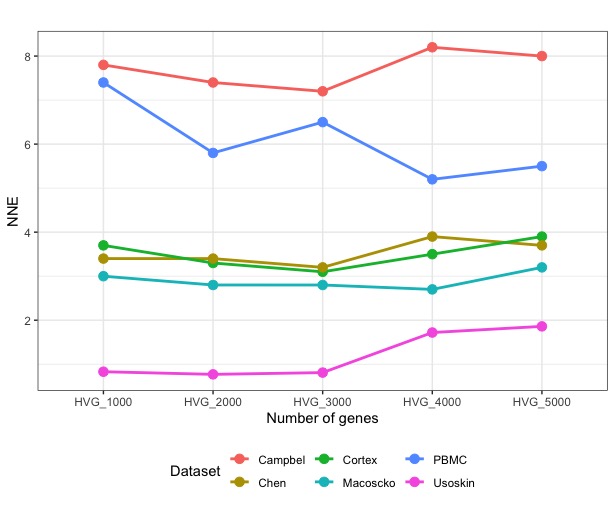}}
\caption{NNE values when different number of high variable genes is used for DAE embedding}
\label{fig7}
\end{figure}

To evaluate the sensitivity of DAE, we performed two sensitivity tests for each dataset. Firstly, we scrutinized the stability of DAE in the context of varying numbers of genes chosen during the gene filtering stage. The results are consolidated in ``Fig.~\ref{fig7}''. As observed, the NNE of DAE remained relatively constant across all datasets.

\begin{figure}[htbp]
\centerline{\includegraphics[width=\linewidth]{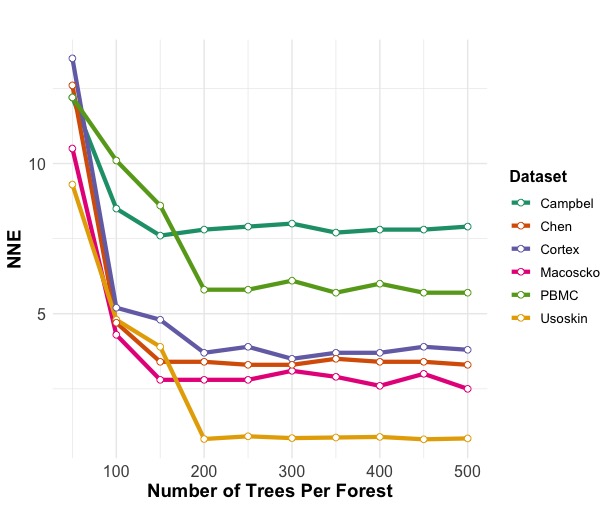}}
\caption{Sensitivity analysis of DAE when various number of trees are used to generated CLG. The performance of DAE reaches a
saturation point when approximately more than 200 trees per forest are utilized.}
\label{fig8}
\end{figure}

Next, we evaluated the influence of varying the number of trees per forest on DAE's performance. This was accomplished by running DAE with a range of tree quantities per forest, from 50 to 500. Following this, we determined the NNE for DAE in terms of 2D data visualization. ``Fig.~\ref{fig8}'' clearly illustrates that the performance of DAE tends to reach a plateau when using roughly more than 200 trees per forest. In this study, all DAE experiments were conducted using 200 trees per forest.

%% Figure 7

%% Figure 8

%% Figure 9
\begin{figure}[htbp]
\centerline{\includegraphics[width=\linewidth]{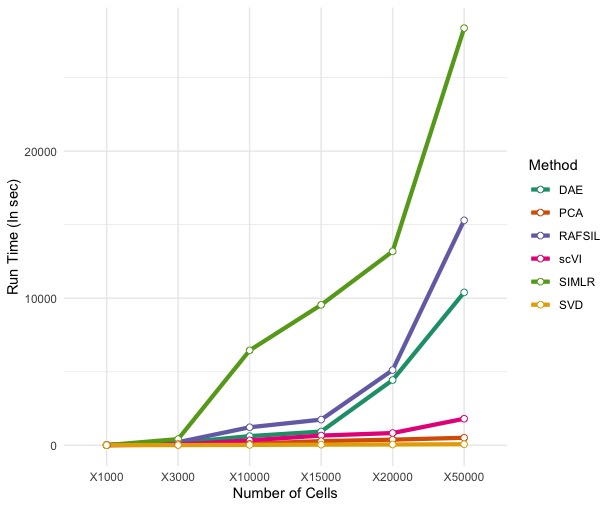}}
\caption{Runtime Analysis}
\label{fig9}
\end{figure}

\subsection{Run time analysis}
\label{sec:sample3.5}
In this part, we compare the running time of our proposed method with other embedding techniques including: RAFSIL, SIMLR, scVI, PCA, and SVD on the PBMC benchmark. For some of the employed techniques (e.g., RAFSIL and SIMLR), we encountered challenges when dealing with the large datasets, as a few methods proved too memory-intensive. As a solution, we utilized a commercial cluster server to generate results for these particular datasets. To handle the computational demands of all methods, we used a high-performance server equipped with an Intel Xeon CPU with 20 physical cores, 256 GB of DDR4 RAM, and 2 TB SSD for fast read/write operations. This setup was used for all methods, experiments, and datasets, allowing us to distribute the computation across multiple cores efficiently and significantly reduce the runtime, making it feasible to apply these techniques to our large-scale single-cell RNA-seq datasets.``Fig.~\ref{fig9}'' provides a breakdown of the average processing time (as per real-world clock time on our reference systems) for each method across the datasets we used. From the data, it's clear that SVD is able to embed data more quickly than their counterparts. Nevertheless, it's worth mentioning that the effectiveness of their embedding could use some enhancement. While DAE's operation speed is tolerable, it's not particularly rapid. However, as pointed out previously, DAE can be conveniently adapted to a parallel mode, which could significantly expedite its speed beyond the present state. This improvement is something we intend to pursue in our future endeavors.

\section{Conclusion}
\label{sec:sample4}

This study, for the first time, unifies gene expression profiles with gene-gene interaction information derived from the same data. While gene expression data provides a direct measure of cellular transcriptional activity, gene-gene interactions offer insights into the regulatory networks that control this activity. By integrating these two complementary data types, our method delivers a more comprehensive and biologically meaningful representation of cellular states, which has the potential to enhance downstream single-cell analyses. As demonstrated through extensive experimental results across six benchmarks, this approach can enhance the quality of single-cell clustering, dimension reduction, and visualization. The positive outcomes of this project encourage us to consider extending the use of our technique to multi-omics single-cell embedding.
While we acknowledge the potential of the method, we recognize that our current investigation may not cover all aspects. While this work provides deep insights into single-cell embedding, the field continues to advance with new methodological developments, and there is ample space for further progress, as is common in newly emerging fields of study. It is important to note that our present research should not be seen as a final conclusion, but rather as a notable step in comprehending cellular complexity, opening up new avenues for research and advancement in the analysis of single-cells.

\end{document}